# *Slow viscoelastic relaxation and aging in aqueous foam*

Sébastien Vincent-Bonnieu, Reinhard Höhler, Sylvie Cohen-Addad

*Université de Marne-la-Vallée*
*Laboratoire de Physique des Matériaux Divisés et des Interfaces, UMR 8108 du CNRS*
*5 Boulevard Descartes, 77 454 Marne-la-Vallée cedex 2, France*

PACS. 83.60.Bc  Linear viscoelasticity
PACS. 83.80.Iz  Emulsions and foams
PACS. 62.20.Hg  Creep

*Like emulsions, pastes and many other forms of soft condensed matter, aqueous foams present slow mechanical relaxations when subjected to a stress too small to induce any plastic flow. To identify the physical origin of this viscoelastic behaviour, we have simulated how dry disordered coarsening 2D foams respond to a small applied stress. We show that the mechanism of long time relaxation is driven by coarsening induced rearrangements of small bubble clusters. These findings are in full agreement with a scaling law previously derived from experimental creep data for 3D foams. Moreover, we find that the temporal statistics of coarsening induced bubble rearrangements are described by a Poisson process.*

A large variety of disordered materials present slow mechanical relaxations when subjected to stresses much too small to induce plastic flow. Glasses undergo slow viscoelastic relaxations with a large spectrum of characteristic times, and their rheology depends on strain history as well as aging. Slow relaxations have also been found in many forms of disordered soft condensed matter such as pastes, emulsions and foams [1,2]. If the constituent particles are small enough for thermal fluctuations to be significant, such materials may be described as "soft glassy materials" [3], in close analogy with conventional glasses. Since in aqueous foams, the constituent particles (the bubbles) are too large for thermal dynamics to be relevant, another source of intrinsic dynamics must be identified. Foams evolve in time due to coarsening, bubble coalescence and drainage of the liquid content [4]. The last two of these mechanisms can be minimized over an extended experimental time scale and will not be considered in this paper. Coarsening results from diffusive gas exchange between neighbouring bubbles, driven by the Laplace pressure which is on the average higher in smaller bubbles than in larger ones. As some bubbles shrink and others grow, their packing can become locally unstable. This leads to intermittent structural rearrangements where small clusters of bubbles switch neighbours so that the topology of the packing changes. The rate of these events which constitute an intrinsic source of dynamics slows down with increasing foam age [5,6]. As another consequence of coarsening, a scaling state is reached in the limit of large foam ages where the temporal evolution of the structure is statistically self similar and the bubble radius increases with foam age following an universal power law [5,7].

Pioneering numerical simulations of dry coarsening foams subjected to step strains [4] as well as rheological experiments with 3D foams [8,9] have shown that there is a strong coupling between the coarsening process and the slow viscoelastic relaxation dynamics of foams. Quantitative insight was obtained in creep experiments with 3D foams where a small shear stress step σ was applied and the strain γ(t) measured as a function of time t. After a brief transient where the response is predominantly elastic, the foam flows like a very viscous Newtonian fluid [9,10]. The rate R of coarsening induced bubble rearrangements per unit time and sample volume was determined in situ by multiple light scattering. Experiments with samples whose coarsening rate varied by more than an order of magnitude yielded the following relation between R, the shear rate $\dot{\gamma}$, the instantaneous elastic shear modulus and a characteristic volume $V$ that scales with foam age as the average bubble volume [9]:

$$\dot{\gamma} = \frac{RV\sigma}{G} \quad (1)$$

To explain this finding, a simple physical picture has been proposed [9]: Small bubble clusters temporarily loose their rigidity as they undergo coarsening induced rearrangements and immediately become rigid again afterwards. During the rearrangement, the average sample elasticity is therefore weakened, as if a liquid-like inclusion were introduced in a homogeneous elastic matrix. As a consequence, the macroscopic strain increases by a small step. This elastic strain increment is transformed into irreversible creep strain when the rearrangement is completed. The macroscopic creep is interpreted as the accumulation of these strain steps. Eq. (1) was derived from such a scenario by a straightforward calculation based on continuum mechanics. However, important questions remain open: Continuum mechanics is adequate for describing behaviour on a macroscopic scale, but using this framework to describe the deformation of rearranging regions which are just a few bubble diameters across is non-trivial. Moreover, individual strain steps associated with rearrangements that should exist according to the conjecture have never been observed experimentally. In a typical 3D sample used in rheological measurements, this would require measuring stress evolutions too small to be resolved by existing rheometers. Moreover, coarsening not only leads to intermittent rearrangements, but there is also a continuous evolution of the structure which might be modified by an applied stress and contribute to the observed creep contribution. The aim of the simulations reported in the present paper is to clarify these issues and to identify the physical origin of quasistatic steady creep in foams on the bubble scale.

Simulating the quasistatic creep rheology of coarsening 3D foams of finite liquid content with enough bubbles to exclude finite size effects and to suppress fluctuations related to the disorder of the samples requires extremely large amounts of computer time. We have therefore chosen to model the behaviour of 2D foams of vanishing liquid content. To create a foam structure, 400 points are randomly dispersed in a unit cell with periodic boundary conditions. The cell is then divided into bubbles by means of the Voronoi construction. The lines separating bubbles are called edges. According to Plateau's rules, the vertices where edges meet must have a coordination of 3 to be stable [4]. Using the Surface Evolver software [11], we relax the structure to minimize the total edge length while conserving the bubble areas. We use a particularly efficient variant of the Surface Evolver software using finite elements that are circular arcs. If upon this calculation an edge length goes to zero, an unstable vertex of coordination larger than 3 is created, leading to a rearrangement of the structure called a T1 event where bubbles switch neighbours. Finally, an equilibrium foam structure is obtained whose second moment of the bubble coordination number distribution is 1.6. To simulate the coarsening of these foams, we use a standard feature of the Surface Evolver allowing diffusive gas transfer between neighbouring bubbles to be implemented. The rate of bubble area evolution due to gas transfer through a given edge is equal to the product of pressure difference, edge length and edge permeability constant. At the same time, the program continually seeks a structure of minimal interfacial energy. We chose a permeability of $10^{-3}$ m Pa$^{-1}$ s$^{-1}$, sufficiently small to achieve continuously good numerical convergence towards a structure of minimal interfacial energy with respect to the instantaneous bubble areas. The average 2D interfacial stress in the foam is evaluated using the well known expression [12]:

$$\sigma_{ij} = \frac{T}{NA} \int (\delta_{ij} - n_i n_j) \, dl \qquad (2)$$

T is the line tension (the 2D analog of surface tension), N the number of bubbles in the sample, A the average bubble area and the integration is carried out over all the bubble edges in the sample. **n** is a unit vector normal to the edge. To apply a controlled stress to the sample as required for a numerical creep experiment, we use a feedback loop: We apply a shear strain that is adjusted whenever the calculated stress deviates from the target value. To shear the sample while maintaining periodic boundary conditions, we apply an affine shear strain to the two basis vectors of the unit cell and an affine displacement to each vertex. The foam structure is then relaxed to find the new configuration of minimal energy.

Using these features, we have performed creep experiments for 10 independently created samples. For each value of the applied stress the simulation required typically one week of CPU time on a PC cluster with 10 processors working in parallel. Since the structures initially created as described above are generally not free of stress, we shear the unit cell in such a way that all stress components are zero. We then apply a fixed constant stress, equilibrate the structure and finally start the coarsening. After 100 s, the coarsening is stopped, the stress reset to zero and the structure again equilibrated. All along this numerical experiment, the average strain in the sample is recorded. Every 0.1 s of simulation time, the structure is checked for configurations requiring T2 processes, defined as the vanishing of bubbles whose size has shrunk to zero, or T1 processes. In both cases, the gas transfer is interrupted and started again only when the structure is equilibrated. In our simulations, most T1 events involve 4 neighbouring bubbles, but we also observe a small number of larger events. The number of bubbles in the sample decreases typically by 14% over the experimental time scale, corresponding to a 7% increase of the average bubble radius. Moreover, to characterize the domain of elastic response of our samples, we have quasistatically increased the applied stress until an equilibrium structure could no longer be established. We identify this maximum stress with the yield stress, denoted $\sigma_y$. Since it is expected to scale as line tension T divided by an average bubble radius <r> [2], we express our result in the dimensionless form $\sigma_y = 0.31$ T/<r>. r is averaged over the entire sample as well as over the duration of the creep experiment. Note that even for stresses well below $\sigma_y$, the rheological response is not always rigorously elastic.

Fig. 1b shows a typical simulated viscoelastic strain of a single foam structure as a function of time, in response to the applied stress shown in fig. 1a. Note that the applied stress remains far below the yield stress. The initial stress step induces an instantaneous deformation, followed by a fluctuating progressive strain increase with time. As the applied stress is finally released, the strain sharply drops by an amount similar to the initial rise, but there is a large residual strain, indicating the presence of irreversible creep flow, in qualitative agreement with the experimental results for 3D foams[9]. A small part of this residual can be explained by plastic flow due to rearrangements triggered as the stress is initially applied at the time 0 s. To show this, we have simulated the strain residual obtained if the stress applied at time 0 s *is released immediately* after its application. The plastic residual obtained in this case, indicated by the grey horizontal line in Fig. 1b, is indeed much smaller than the one obtained after the creep flow. Note that at t = 0 s, the elastic deformation can be calculated by subtracting the plastic residual from the instantaneous deformation. Moreover, during the creep experiment, part of the strain increase is due to coarsening induced softening. By averaging over 10 foam structures, we obtain a relative increase of 0.6% of the elastic deformation between times t = 0 s and t = 100 s, which is much smaller than the irreversible creep. In contrast to the 3D experimental results, the simulated creep does not increase linearly with time. This is so because contrary to the studied 3D foams, the simulated foams are initially far from a scaling state, and it takes about 40 s of coarsening for T1 events to set in. This is shown in fig. 1c which also indicates that the increase of the T1 rearrangement rate is proportional to the increase of the creep rate, in full agreement with the empirical law eq. (1) derived from the 3D experiments. This result is particularly significant because it shows that the prediction of eq.(1) does not only hold under steady state conditions or in the scaling state. To study the linearity of the creep response with applied stress, we plot in Fig. 1c the time evolution of the compliance, defined as the strain divided by the applied stress. The data are averaged of 10 samples to reduce fluctuations. There is good agreement among the compliances obtained for different values of the applied stress, indicating that the response is indeed linear, as predicted by eq. (1).

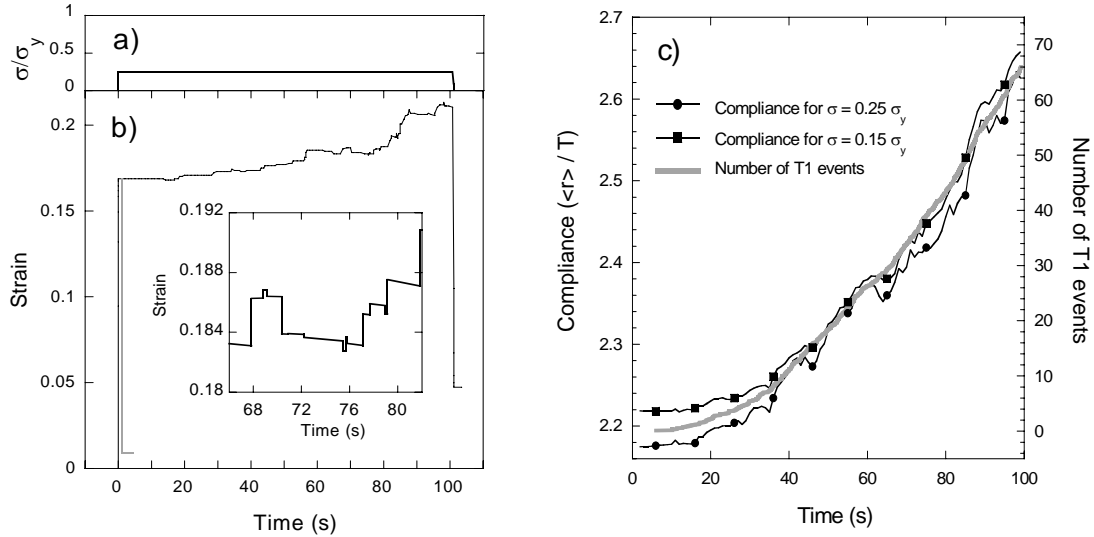

**Fig. 1-** *a) Applied stress $\sigma$, normalized by the yield stress $\sigma_y$. b) Typical strain response of a single coarsening sample. The grey line at the time 0 s indicates the residual plastic strain obtained if stress is applied as illustrated in 1a), but released immediately. The inset of the figure is a detailed view of the strain evolution. c) Temporal evolution of the compliance, normalized by $<r>/T$ and averaged over 10 samples. The grey line represents the total number of T1 events detected since the stress has been applied and coarsening has started. These latter data are the same for the two applied stresses.*

The results discussed up to now show that our simulations reproduce and extend the range of validity of the scaling law eq.(1) governing the creep flow of real 3D foams. To get deeper insight, we analyze how creep flow arises on the bubble scale. The inset of Fig. 1b is an expanded view of a typical temporal strain evolution in one of the 10 foam samples during the creep flow. We can distinguish two kinds of evolution with time: Abrupt steps, alternating with roughly linear continuous variations between successive steps. Extensive simulations have shown that the *strain steps systematically coincide with coarsening induced T1 events*. T2 events were found not to have any direct influence on macroscopic strain. We consider the continuous and the abrupt strain variations during the creep flow as two distinct random variables and calculate their respective average and standard deviation as a function of the applied stress. Despite of the strong fluctuations, the large number of analyzed variations allows the respective average values to be determined very accurately. Assuming Gaussian statistics, the expected error of the mean values corresponds to the symbol sizes used in the plot. Fig. 2 clearly confirms a previous conjecture [9] according to which the average step size should scale linearly with applied stress.

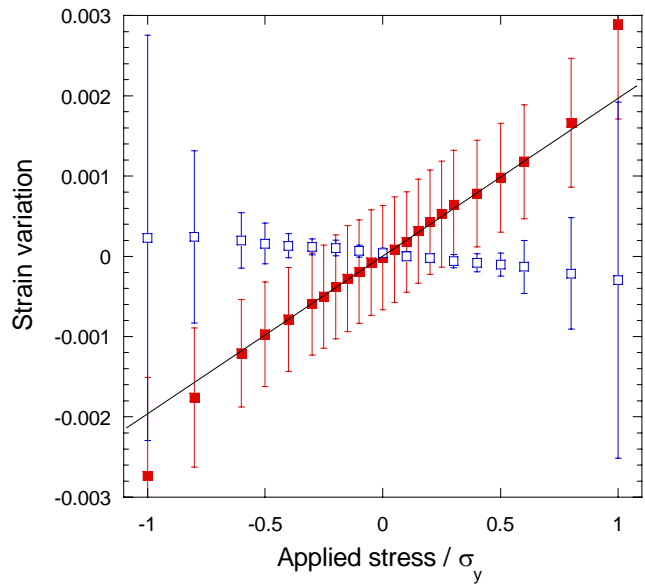

***Fig. 2 -*** *Variation of strain accompanying a T1 event (full symbols) or between two successive T1 events (empty symbols) plotted as a function of the applied stress. The error bars show the standard deviations. The statistical error of the respective mean values is of the order of the symbol size. The straight line is a linear fit to the data for stresses ranging from -0.5 $\sigma_y$ to 0.5 $\sigma_y$.*

Moreover, the average creep due to the strain steps upon T1 events dominates by an order of magnitude over the creep due to the continuous strain evolution between successive T1's. To distinguish elastic from plastic contributions, we have determined the shear modulus of the foam structures just before and just after each T1 event. A T1 induces principally a plastic irreversible strain accompanied by a smaller elastic strain. The latter is due to softening of the foam, related to the T1 induced decrease of edge length density. The strain evolution between successive T1's is also due to elastic as well as plastic contributions, leading to the small negative strain variation shown in fig. 2. These results will be presented in detail elsewhere[13].

To gain insight about the plastic strain variation between successive T1's, we have studied 2D dry foams containing mainly hexagonal bubbles in which 4, 5 or 7 sided bubbles are introduced as "defects". These samples are constructed so that no T1 or T2 rearrangements occur even if coarsening strongly changes the bubble sizes in the vicinity of the defects. After a creep experiment of the type defined in Fig. 1a, these foams do present some residual strain but, contrary to the experiments discussed above, it is independent of the stress applied during the coarsening. This result may be related to Von Neumann's law according to which the growth of each bubble in a dry 2D foam is exclusively determined by its number of neighbours and the edge permeability[4,14]. Therefore, as a consequence of the constant topology, the time evolution of each bubble area in the sample is independent of the applied stress. The independence of the residual strain on applied stress suggests that bubble area and bubble topology uniquely fix the stress free reference state obtained when the stress is finally released, recalling a proposed uniqueness conjecture for dry 2D foams[15]. Recently, 2D bubble clusters have been constructed where several distinct states of equilibrium exist for a given bubble area and topology distribution[15,16], but if such states exist in 2D bulk foam, they do not appear to be relevant in our creep experiments. Let us note that in our simulations over periods of time sufficiently long for T1 events to occur, the structure and topology changes resulting from these events were in some cases found to depend on the applied stress. We expect this feature to be crucial for explaining the stress dependence of the continuous creep flow between successive rearrangements.

We now discuss the dominant creep mechanism, due to the strain steps upon T1 events. Their average size varies linearly with the applied stress almost up to the yield stress, in agreement with the continuum mechanics model cited above. Moreover, according to eq.(1) the size of individual strain steps multiplied by $G/\sigma$ represents the effective fraction of the sample that is predicted in this framework to loose its rigidity upon a rearrangement. The data shown in fig. 2 yield a fraction of 0.003, corresponding for our samples to an effective rearrangement size of the order of 1 bubble. Most T1 events in our 2D simulation involve 4 neighbouring bubbles, which is significantly larger than this value. Rearranging bubbles are therefore much less efficient in relaxing strain than predicted by a simple continuum mechanics picture. Future more refined models will have to take into account explicitly the structure of foams on the bubble scale. Let us finally note that the effective volume fraction of a rearrangement determined from 3D experiments with wet foams is an order of magnitude larger than the value obtained here for dry 2D foam. This may be so because rearrangements in 3D tend to involve intrinsically more bubbles than in 2D and also because the size of rearrangements is an increasing function of foam liquid fraction [4].

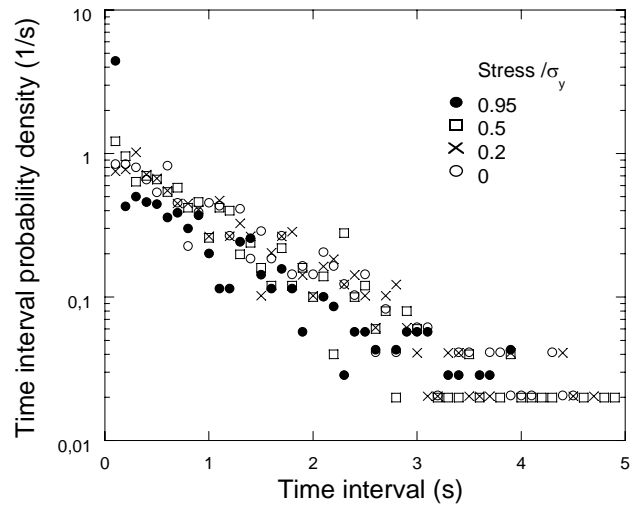

*Fig. 3 - Probability distribution of time intervals between successive T1 events in the foam sample for different constant applied stresses in a range going up to the yield stress. The bin size used to calculate these distributions is 0.1 s.*

We have also studied the distribution of time intervals between successive T1's during the creep simulations, as a function of applied stress. These data, shown in Fig. 3, are based on events at simulation times > 40 s where the rearrangement rate is roughly steady (cf.Fig. 1c). Results for time intervals above 5 s are not displayed since they are very noisy. For all stresses $\leq 0.5\ \sigma_y$, the time interval distributions decrease exponentially with interval duration. This result, expected for a Poisson process, suggests that successive events are statistically independent in this regime. Therefore, the slope of the exponential decrease, represented in the semi-logarithmic graph, is the inverse of the average time between rearrangements which does not depend on the applied stress. Manifestly, coarsening dynamics and creep are governed by this single characteristic timescale. Moreover, we conclude that coarsening dynamics are robust with respect to changes of elastic stress. For the highest investigated stress which is very close to the yield stress, the sharp peak at very small time intervals indicates avalanche-like rearrangement sequences where one event helps to trigger the following one. They may be precursors of shear banding. Studying the transition from creep to yielding is therefore an interesting perspective for further work.

The Surface Evolver simulations presented in this paper show that slow linear viscoelastic dynamics in 2D dry coarsening foams fundamentally differ from those in other disordered materials such as soft glasses. A single characteristic time scale, given by the average time interval between coarsening induced bubble rearrangements, governs the viscoelastic long time response. It is mainly due to intermittent T1 rearrangements of small bubble clusters. They lead to abrupt macroscopic strain steps that are biased by the applied stress. Moreover, the simulations are fully consistent with the scaling law governing creep flow that had previously been observed in 3D experiments. The work presented here provides new insight about the origin of slow relaxations in foam. This will be useful for developing more generic models of the interplay between macroscopic rheology and local structural rearrangements in similar forms of soft condensed matter.

*Acknowledgements* - We thank A. Kraynik, D. Weaire, K. Brakke, S. Cox and the participants of the Foam Rheology in Two Dimensions Workshop held in Aberystwyth in 2005 for stimulating discussions.